\documentclass[preprint]{jpsj2}

\newcommand{\etal}{\textit{et al. \ }}

\title{Calculation of Penetration Depth and $T_c$ in $\kappa $-$($ET$)_2$Cu(NCS)$ _2$ under Pressure}

\author{Kazunori \textsc{Tanaka} \thanks{E-mail address: tanaka. kazunori@scphys.kyoto-u.ac.jp},    Hiroaki \textsc{Ikeda} and Kosaku \textsc{Yamada}}

\inst{Department of science,    Kyoto university,    Kyoto 606-8502,    Japan}
\recdate{July 7, 2004}

\abst{ The pressure dependence of the inverse square of the magnetic penetration depth $\lambda ^{ - 2}$ in  $\kappa$-$($ET$)_2$Cu(NCS)$ _2$ was measured by Larkin, \etal According to the paper, $\lambda ^{ - 2}$ behaves differently under low pressure and under high pressure. Under low pressure, the development of $\lambda ^{ - 2}$ just below $T=T_c$ is rapid compared to the case under high pressure. Moreover, $T_c$ in $\kappa$-$($ET$)_2$Cu(NCS)$ _2$ increases under $c$-axis pressure up to 1kbar and decreases under higher pressure, while $T_c$ decreases monotonically under the hydrostatic pressure, or under the uniaxial pressure parallel to other axes.  In order to explain these behaviors, we calculate $T_c$ and $\lambda ^{ - 2}$ for  $\kappa$-$($ET$)_2$Cu(NCS)$ _2$ under pressure. In the calculation we mainly use an effective dimer Hubbard model. In conclusion, the behavior of $\lambda ^{ - 2}$ results from three effects: the variation of the bandwidth of quasiparticles, the change of the Fermi surfaces, and the effect of vertex correction. This is a different mechanism from that of  $\lambda ^{ - 2}$ in cuprates which we observe when the doping varies. Moreover, we explain the increase in  $T_c$ under the $c$-axis pressure up to 1kbar and the decrease in $T_c$ over 1kbar from our calculation. With the increase in the $c$-axis pressure, two competitive effects with respect to $T_c$ appear. One is the approach of the Fermi surface to the antiferromagnetic Brillouin zone boundary, and the other is the suppression of the electron correlation. Under the low $c$-axis pressure, $T_c$ increases since the former effect is dominant. On the other hand, $T_c$ decreases since the latter effect is dominant under the high $c$-axis pressure.}

\kword{unconventional superconductivity, organic superconductor, penetration depth, vertex correction, pressure}

\begin{document}
\maketitle

\section{\label{sec0a}Introduction}
Recently, many measurements have been performed in the superconducting state in organic conductors including $\kappa$-$($ET$)_2$Cu(NCS)$ _2$. Many interesting behaviors are observed under pressure in organic conductors, since they are soft materials. Among them, we study the pressure dependence of the inverse square of the magnetic penetration depth $\lambda ^{ - 2}$ and the behavior of the superconducting transition temperature $T_c$ under the hydrostatic and the uniaxial pressure by using the effective dimer Hubbard model.

We calculate $\lambda ^{ - 2}$ in $\kappa$-$($ET$)_2$Cu(NCS)$ _2$ on the basis of Jujo's theory. \cite{bib1,bib1a} This theory is a general theory on the transport phenomena and the magnetic penetration depth in the superconducting state, including the effect of the vertex correction, and can be applied to unconventional superconductors such as cuprates and organic superconductors $\kappa$-$($ET$)_2$X. By taking account of the vertex correction, which can be considered as the backflow effect, this theory explains why the value of inverse square of the magnetic penetration depth $ \lambda^{-2}$ is suppressed near $T=0$ under the strong antiferromagnetic fluctuation in unconventional superconductors, such as cuprates in the underdoped region. In this paper we discuss the pressure dependence of the extrapolated value of $ \lambda^{-2}$ at $T=0$ and the development of $ \lambda^{-2}$ just below $T_c$. Moreover, $T_c$ under pressure is calculated by the linearized Dyson-Gor'kov equations, and the different dependence of $T_c$ on hydrostatic and uniaxial pressure is explained.

\section{\label{sec3} The Pressure Dependence of $\lambda^{-2}$ and of $T_c$ in $\kappa$-$($ET$)_2$Cu(NCS)$ _2$ }
The pressure dependence of the inverse square of the magnetic penetration depth $\lambda^{-2}$ in $\kappa$-$(d_8$ET$)_2$Cu(NCS)$ _2$ was measured by Larkin, \etal \cite{bib4} Here, we refer to $\kappa$-$($ET$)_2$Cu(NCS)$ _2$ in which hydrogens in ET molecules are deuterated as $\kappa$-$(d_8$ET$)_2$Cu(NCS)$ _2$, while the materials with protonated ones are referred to as $\kappa$-$(h_8$ET$)_2$Cu(NCS)$ _2$.  Ref.\citen{bib4} shows that the value of $\left| {\frac{{d\lambda ^{ - 2} }}
{{dT}}} \right|_{T = T_c } $ is large under low pressure ($P=0$bar) compared to that under high pressure($P$ is up to 1290 bar). This implies that there is a rapid development of $\lambda^{-2}$ just below $T_c$ under low pressure. Moreover, Ref.\citen{bib4} shows that $\lambda^{-2}$ is suppressed in the low temperature region under low pressure, although the data near $T=0$ is insufficient.  This suppression is analogous to the doping dependence of $\lambda^{-2}$ in high-$T_c$ cuprates such as LSCO\cite{bib5}. The $\lambda^{-2}$-$T$ diagram in the underdoped and overdoped LSCO behaves similarly to  $\kappa$-$($ET$)_2$Cu(NCS)$ _2$ under low pressure and under high pressure, respectively.  The behavior of $\lambda^{-2}$ in cuprates is understood from the fact that the effect of the vertex correction is strong in the underdoped region, and weak in the overdoped region. The vertex correction affects strongly under low pressure and weakly under high pressure in $\kappa$-$($ET$)_2$Cu(NCS)$ _2$. In this paper, we study the behavior of  $\lambda^{-2}$ in $\kappa$-$($ET$)_2$Cu(NCS)$ _2$ theoretically by carrying out the actual calculation.

Next, we refer to the pressure dependence of $T_c$ for $\kappa $-$($ET$)_2$Cu(NCS)$ _2$. $T_c$ under the hydrostatic pressure have been measured by many experiments.\cite{bib7a,bib7b,bib11} According to the results, $T_c$ decreases monotonically as the pressure increases. On the other hand, the behavior of $T_c$ under the uniaxial pressure is quite different from that under the hydrostatic pressure according to the recent measurements for $\kappa$-$($ET$)_2$Cu(NCS)$ _2$\cite{bib18, bib19, bib20}. When the uniaxial pressure is applied in parallel to $b$-axis or to $a$-axis, $T_c$ decreases. This is similar to the behavior under the hydrostatic pressure. On the other hand, $T_c$ increases up to 1kbar under the $c$-axis pressure and decreases under higher pressure, which is in contrast to the behavior of $T_c$ under the hydrostatic pressure. By calculation, we show that this is because two competitive effects with respect to $T_c$ appear under the $c$-axis pressure: the approach of the Fermi surface to the antiferromagnetic Brillouin zone boundary and the suppression of the electron correlation.

\section{\label{sec1a} Effective Dimer Hubbard model for Organic Superconductor $\kappa$-$($ET$)_2$X }
\begin{figure}
\includegraphics[width=7.5cm]{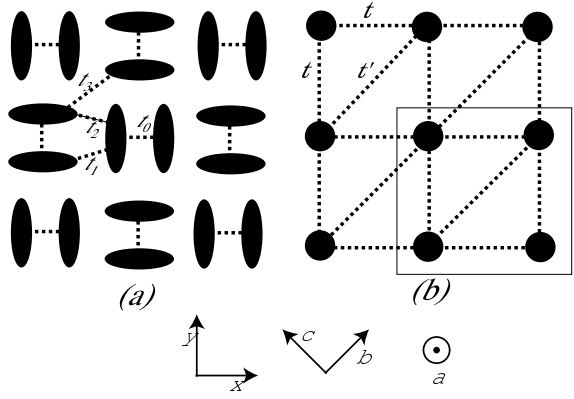}
\caption[]{(a): The original lattice structure of $\kappa$-$($ET$)_2$X, where the ellipses represent ET molecules. A dimer consists of two parallel ET molecules.
(b): The lattice structure of $\kappa$-$($ET$)_2$X when the pair of ET molecules are dimerized.\cite{bib6, bib7} The circles represent dimers, and the rectangle represents the unit cell. $\kappa$-$($ET$)_2$X is quasi-two-dimensional (Q2D) conductor where the layer is usually denoted as the $bc$-plane. $x$- and $y$-axes are introduced for the convenience of calculation. $a$-axis is perpendicular to the $bc$-plane.}
\label{fig000}
\end{figure}

We use the effective dimer Hubbard model for $\kappa$-$($ET$)_2$X.\cite{bib6, bib7, bib7a, bib7b} In this paper we adopt a quasi-two-dimensional (Q2D) band structure ignoring the interlayer hopping term for $\kappa$-$($ET$)_2$X and symmetrize two dimers in the unit cell\cite{bib7b}. Then the Hamiltonian is expressed as follows\cite{biborg}:
%
%
\begin{align}
H = \sum\limits_{\mib{k}} {\varepsilon _{\mib{k}} c_{\mib{k}\sigma } ^\dag  c_{{\mib{k}}\sigma } }  + U\sum\limits_{{\mib{k,k',q}}} {c_{\mib{k}\uparrow }  ^\dag  c_{{\mib{q - k}}\downarrow  }  ^\dag  c_{{\mib{q - k'}}\downarrow }   c_{{\mib{k'}} \uparrow } } ,\label{eqH1}\\
\varepsilon _{\mib{k}}  =  - 2t \left( {\cos k_x  + \cos k_y } \right) - 2t' \cos (k_x  + k_y )  , \label{eqH2}
\end{align}
where $t$ and $t'$ are the nearest-neighbor and next-nearest-neighbor hopping terms shown in Fig.\ref{fig000}. $U$ is the Coulomb repulsion and $\varepsilon _{\mib{k}}$ is the dispersion. The bandwidth $W$ is expressed as the maximum of the dispersion measured from the minimum: $\varepsilon _{\mib{k}max}-\varepsilon _{\mib{k}min} \sim 8t$. The exact value of $W$ is $8.22t$ and $8.39t$ at $t'/t=0.7$ and $0.75$, respectively. In $\kappa$-$($ET$)_2$X, the electrons are half-filled.
%
%
%
%
%
In this paper, we adopt the FLEX approximation, the self-consistent second order perturbation theory (SC-SOPT), and the third order perturbation theory (TOPT) in the calculation of the self-energy terms.\cite{bib1,bib1a} We adopt different approximations in the calculation for $\lambda^{-2}$ and for $T_c$. In the calculation for $\lambda^{-2}$ FLEX and SC-SOPT are adopted, since conservation approximations are needed. FLEX is appropriate for the case with the strong antiferromagnetic fluctuation, while SC-SOPT is used to show the weak correlation effect and the weak antiferromagnetic fluctuation qualitatively. On the other hand, SC-SOPT is not enough to give actual $T_c$ since the third order term in $U$ is important for calculating $T_c$. Therefore when the antiferromagnetic fluctuation is weak, we adopt TOPT instead of SC-SOPT in the calculation for $T_c$. As a conclusion, we adopt FLEX and TOPT in the calculation for $T_c$. The former is used for the strong antiferromagnetic fluctuation and the latter is used for the weak correlation.

Within FLEX and SC-SOPT, the self-energy term and the Green's function in the normal state $\Sigma_n$ is expressed as follows:
%
%
\begin{align}
\Sigma _n \left( k \right) = \frac{T}
{N}\sum\limits_q {V\left( q \right)G\left( {k - q} \right)},  \label{eqsig1} \\
G\left( k \right) = \frac{1}{{\mathrm{i} \varepsilon _n  - \left( {\varepsilon _{\mib{k}}  - \mu } \right) - \Sigma _n \left( k \right)}}  , \label{eqsig11}
\end{align}
%
%
where $k = \left( {{\mib{k}},\varepsilon _n } \right)$, $\varepsilon _n  = \left( {2n + 1} \right)\pi T$ is the fermion Matsubara frequency, and $\mu$ is the chemical potential. The normal vertex $ V\left( q \right)$ in Eq.\ref{eqsig1} is given by
%
\begin{align}
V\left( q \right) = \frac{3}
{2}U^2 \frac{{\chi \left( q \right)}}
{{1 - U\chi \left( q \right)}} + \frac{1}
{2}U^2 \frac{{\chi \left( q \right)}}
{{1 + U\chi \left( q \right)}} - U^2 \chi \left( q \right)  \label{eqs3}
\end{align}
%
for FLEX and 
%
%
\begin{align}
V\left( q \right) =U^2 \chi \left( q \right) \label{eqs4}
\end{align}
%
%
for SC-SOPT. The spin susceptibility $\chi \left( q \right)$ is expressed as 
%
%
\begin{align}
\chi \left( q \right) =  - \frac{T}{N}\sum\limits_{q} {G\left( k+q \right)G\left( k \right)}. \label{eq202a}
\end{align}
%
Note that $q = \left( {{\mib{q}},\omega _l } \right)$, and $\omega _l  = 2n\pi T$ is the boson Matsubara frequency. Within FLEX and SC-SOPT, we obtain the normal self-energy term and the normal Green's function by solving Eqs.\ref{eqsig1} and \ref{eqsig11} self-consistently.

Within TOPT, the normal self-energy term is expressed as follows.
%
%
\begin{align}
\Sigma _n \left( k \right) = \frac{T}{N}\sum\limits_{k'} {\left[ {U^2 \chi _0 \left( {k - k'} \right) + U^3 \chi _0 ^2 \left( {k - k'} \right) + U^3 \phi _0 ^2 \left( {k + k'} \right)} \right]G_0 \left( {k'} \right)} , \label{topt1} \\
\chi _{0} \left( q \right) =  - \frac{T}{N}\sum\limits_k {G_0 \left( k \right)} G_0 \left( {k + q} \right),     \label{topt2} \\
\phi _{0} \left( q \right) =  - \frac{T}{N}\sum\limits_k {G_0 \left( k \right)} G_0 \left( {q - k} \right),  \label{topt3} 
\end{align}
%
%
where $\chi _{0} \left( q \right)$ and  $\phi _{0} \left( q \right)$ are calculated from the bare Green's function $G_0  \left( k \right) =\frac{1}{{\mathrm{i} \varepsilon _n  - \left( {\varepsilon _{\mib{k}}  - \mu } \right) }}$ and $\chi _{0} \left( q \right) \ne \chi \left( q \right)$. This shows that the calculation of the self-energy term and the Green function within TOPT is not self-consistent in contrast to FLEX and SC-SOPT. This is because self-consistent TOPT seems to be complicated in actual calculation.

Here we introduce the FLEX approximation in the superconducting state, the normal and anomalous self-energy terms $\Sigma _{n} \left( k \right) $ and $\Sigma _{a} \left( k \right) $ are given by 
%
\begin{align}
\Sigma _{n} \left( k \right) = \frac{T}
{N}\sum\limits_q {V_{n}\left( q \right)G\left( {k - q} \right)}, \label{se1} \\
\Sigma _{a} \left( k \right) =- \frac{T}
{N}\sum\limits_q {V_{a}\left( q \right)F \left( {k - q} \right)}, \label{se2}
\end{align}
%
%
where $ V_{n}\left( q \right)$ and $ V_{a}\left( q \right)$ are the normal and anomalous vertices in the superconducting state, respectively. 
The normal and anomalous Green's functions $ G\left( {k - q} \right)$ and $ F\left( {k - q} \right)$ are written as 
%
%
\begin{align}
G\left( k \right) = \frac{{{\rm{i}}\varepsilon _n  + \left( {\varepsilon _{\mib{k}}  - \mu } \right) + \Sigma _n \left( { - k} \right)}}{{\left[ {{\rm{i}}\varepsilon _n  - \left( {\varepsilon _{\mib{k}}  - \mu } \right) - \Sigma _n \left( k \right)} \right]\left[ {{\rm{i}}\varepsilon _n  + \left( {\varepsilon _{\mib{k}}  - \mu } \right) + \Sigma _n \left( { - k} \right)} \right] - \Sigma _a \left( k \right)^2 }}, \label{scg1} \\
F\left( k \right) = \frac{{ - \Sigma _a \left( k \right)}}{{\left[ {{\rm{i}}\varepsilon _n  - \left( {\varepsilon _{\mib{k}}  - \mu } \right) - \Sigma _n \left( k \right)} \right]\left[ {{\rm{i}}\varepsilon _n  + \left( {\varepsilon _{\mib{k}}  - \mu } \right) + \Sigma _n \left( { - k} \right)} \right] - \Sigma _a \left( k \right)^2 }}. \label{scg2}
\end{align}
%
%
$ V_{n}\left( q \right)$ and $ V_{a}\left( q \right)$ are given by
%
%
\begin{align}
V_{n} \left( q \right) = \frac{{U^2 }}{2}\left[ {3\frac{{\chi _s \left( q \right) }}{{1 - U\chi _s \left( q \right) }} - \chi _s  \left( q \right) + \frac{{\chi _c  \left( q \right)}}{{1 + U\chi _c \left( q \right) }} - \chi _c \left( q \right) } \right], \label{scv1} \\
V_{a} \left( q \right) = \frac{{U^2 }}{2}\left[ {3\frac{{\chi _s \left( q \right) }}{{1 - U\chi _s \left( q \right) }} - \chi _s \left( q \right)  - \frac{{\chi _c \left( q \right) }}{{1 + U\chi _c \left( q \right) }} + \chi _c \left( q \right) } \right], \label{scv2}
\end{align}
%
%
with 
%
%
\begin{align}
\chi _{\scriptstyle s \hfill \atop 
  \scriptstyle c \hfill} \left( q \right) =  - \frac{T}{N}\sum\limits_q {\left[ {G\left( {k + q} \right)G\left( k \right) \pm F\left( {k + q} \right)F\left( k \right)} \right]}, \label{eqkai}
\end{align}
%
%
where $+$ and $-$ correspond to $\chi _s$ and $\chi _c$, respectively.

We solve Eqs.\ref{se1}-\ref{eqkai} self-consistently to calculate $ G\left( k \right)$, $ F\left( k \right)$, $\Sigma _{n} \left( k \right)$ and $\Sigma _{a} \left( k \right)$. The symmetry of Cooper pair in $\kappa$-$($ET$)_2$X  is $d_{x^2-y^2}$, with which the superconducting gap $\Sigma _a \left( k \right)$ has nodes in $ \pm \pi/4$ directions in the $k$-space.\cite{biborg} The $d_{x^2-y^2}$ gap symmetry for $\kappa$-$($ET$)_2$X is justified later in calculating $T_c$. In order to realize this symmetry, we set $\Sigma _a \left( \mib{k}, \varepsilon _{ \pm 1}  \right) \propto \left( {\cos k_x  - \cos k_y } \right)$ as the intial value for solving Eqs.\ref{se1}-\ref{eqkai}.

\section{\label{sec1b} Calculation of $T_c$ by Dyson-Gor'kov Equations}
In this section, we calculate $T_c$ in $\kappa$-$($ET$)_2$X by the linearized Dyson-Gor'kov equation within FLEX\cite{bib14, bib15, bib16, bibr1, bibr2} and TOPT.\cite{biborg} First of all, the normal and anomalous Green's functions satisfy Dyson-Gor'kov equations.\cite{bib13}
\begin{align}
G \left( k \right) = G_0 \left( k \right) + G_0 \left( k \right)\Sigma _n \left( k \right)G \left( k \right) + G_0 \left( k \right)\Sigma _a \left( k \right)F^\dag  \left( k \right) \label{eq5106}, \\
F^\dag  \left( k \right) = G_0 \left( -k \right)\Sigma _n \left(- k \right)F^\dag \left( k \right) + G_0 \left( -k \right)\Sigma _a \left( -k \right)G \left( {k} \right) \label{eq5107}. 
\end{align}
When $T$ approaches $T_c$ within the superconducting state, Eqs.\ref{eq5106} and \ref{eq5107} are linearized as follows, since $ F\left( k \right)  \ll G\left( k \right)$.
\begin{align}
F\left( k \right) = \left| {G \left( k \right)} \right|^2 \Sigma _a \left( k \right),  \label{eq5108} \\
G \left( k \right) = G_0 \left( k \right) + G_0 \left( k \right)\Sigma _n \left( k \right)G \left( k \right) \label{eq5109}.
\end{align} 
Within FLEX, $\Sigma _n \left( k \right)$ and $\Sigma _a \left( k \right)$ in the vicinity of $T_c$ are calculated by linearizing Eqs.\ref{se1} and \ref{se2}, respectively. Then, $\Sigma _n \left( k \right)$ is the same as that in the normal state, which is expressed by Eqs.\ref{eqsig1} and \ref{eqs3}, while $\Sigma _a \left( k \right)$ is expressed by the linearized Dyson-Gor'kov equation:
\begin{align}
\Sigma _a \left( k \right) =  - \frac{T}{N}\sum\limits_{k'} {V_a \left(q \right)\left| {G \left( {k-q} \right)} \right|} ^2 \Sigma _a \left( {k-q} \right), \label{eq6103} \\
V_a \left( q \right) = U^2 \left[ {\frac{3}{2}\frac{\chi\left( q \right) }{{1 - \chi \left( q \right)}} - \frac{1}{2}\frac{\chi\left( q \right) }{{1 + \chi \left( q \right)}}} \right] + U. \label{eq6104}
\end{align}

On the other hand, the linearized Dyson-Gor'kov equation for TOPT is 
\begin{align}
\Sigma _a \left( k \right) =  - \frac{T}{N}\sum\limits_{k'} {V_a \left( {k, k'} \right)\left| {G \left( {k'} \right)} \right|} ^2 \Sigma _a \left( {k'} \right), \label{topt11}
\end{align}
where,  
\begin{align}
V_a \left( {k, k'} \right) = V_{RPA} \left( {k, k'} \right)+ V_{vert} \left( {k, k'} \right) , \label{topt12a}\\
V_{RPA} \left( {k, k'} \right)= U + U^2 \chi _0 \left( {k + k'} \right) + 2U^3 \chi _0 ^2 \left( {k + k'} \right) , \label{topt12b}\\
V_{vert} \left( {k, k'} \right)=2U^3 \frac{T}{N}{\mathop{\rm Re}\nolimits} \sum\limits_{k_1 } {G_0 \left( {k_1 } \right)G_0 \left( {k + k_1  - k'} \right)\left[ {\chi _0 \left( {k + k_1 } \right) - \phi _0 \left( {k + k_1 } \right)} \right]}. \label{topt12c}
\end{align}

If we replace the left hand side of Eqs.\ref{eq6103} and \ref{topt11} by $ \alpha \Sigma _a \left( k \right)$, this equation can be considered as an eigenvalue equation with eigenvalue $\alpha$ and eigenvector $\Sigma _a \left( k \right)$. $\mathit{T_{c}}$  is the temperature at which the maximum eigenvalue  $\alpha_{max}$ reaches to unity. $\Sigma _a \left( k \right)$ with which the largest eigenvalue is obtained represents the superconducting gap symmetry. Among several gap functions, the $d_\mathrm {x^{2}-y^{2}}$ state possesses the maximum eigenvalue in the region $0.17<t/U<0.35$ and $0.4<t'/t<0.8$ within FLEX, and $0.14<t/U<0.25$ and $0.4<t'/t<0.8$ within TOPT.

\section{\label{sec2} Effect of the vertex correction and $\lambda^{-2}$ : Introduction to Jujo's theory. }
In this section, we introduce Jujo's theory shortly. From Kubo formula, the penetration depth is expressed with the electromagnetic response kernel as follows. 
%
%
\begin{align}
\lambda _{\mu \nu } ^{ - 2}  =  - 4\pi K_{\mu \nu } \left( {q \to 0} \right), \label{eq001} 
\end{align}
%
\begin{align}
K_{\mu \nu}  \left( {q \to 0} \right) = -\frac{T}
{N}\sum\limits_{{\mib{k}},n} {{\text{Tr}}} \left[ {\hat \Lambda _\mu  ^0 \left( {k,k+q} \right)\hat G\left( k+q \right)\hat \Lambda _\nu  \left( {k+q,k} \right)\hat G\left( k \right)} \right]_{q \to 0} - \frac{T}
{V}\sum\limits_{{\mib{k}},n} {\frac{{\partial ^2 \varepsilon _{\mib{k}}^{} }}
{{\partial k_\nu  \partial k_\mu  }}} \operatorname{Tr} \left[ {\hat \tau _3 \hat G\left( k \right)} \right]e^{i\varepsilon _n 0_ +  }. \label{eq201}
\end{align}
%
Equation \ref{eq201} is expressed by Nambu matrices. The Green's function matrix is defined as:
%
\begin{align}
\hat G\left( k \right) = \left( {\begin{array}{*{20}c}
   {G\left( k \right)} & {F\left( k \right)}  \\
   {F\left( k \right)} & { - G\left( -k \right)}  \\
\end{array}} \right)  \nonumber \\
{} = \frac{1}{{{\rm{i}}\varepsilon _n \hat \tau _0  - \varepsilon _{\mib{k}} \hat \tau _3  - \hat \Sigma \left( k \right)}}, \label{eqnam1}
\end{align}
%
where the self-energy matrix $\hat \Sigma \left( k \right)$ is given by
%
%
\begin{align}
\hat \Sigma \left( k \right) = \left( {\begin{array}{*{20}c}
   {\Sigma _n \left( k \right)} & {\Sigma _a \left( k \right)}  \\
   {\Sigma _a \left( k \right)} & { - \Sigma _n \left( { - k} \right)}  \\
\end{array}} \right). \label{eqnamx}
\end{align}
%
%
Matrices $\hat \tau _0 $ and $\hat \tau _3 $ are given by
%
%
\begin{align}
 \hat \tau _0  = \left( {\begin{array}{*{20}c}
   1 & 0  \\
   0 & 1  \\
\end{array}} \right), \label{eqnam2} \\
\hat \tau _3  = \left( {\begin{array}{*{20}c}
   1 & 0  \\
   0 & { - 1}  \\
\end{array}} \right) . \label{eqnam3}
\end{align}
%
\begin{align}
\left. {\hat \Lambda _\mu ^0 \left( {k,k + q} \right)} \right|_{q \to 0}  = \left( {\begin{array}{*{20}c}
   {v_{{\mib{k}}\mu } } & 0  \\
   0 & {v_{{\mib{k}}\mu } }  \\
\end{array}} \right) \label{eqnam4}
\end{align}
%
is the bare three-point vertex, where $v_{\mib{k}}  = \frac{{\partial \varepsilon _{\mib{k}} }}{{\partial {\mib{k}}}}$ is the bare velocity of electrons. $
\hat \Lambda _\mu  \left( {k + q,k} \right)$ satisfy the following integral equation:
%
%
\begin{align}
\hat \Lambda _{\mu \;i,j} \left( {k + q,k} \right) = \hat \Lambda _{\mu \;i,j}^0 \left( {k + q,k} \right) \nonumber \\
{}  + \frac{T}{N}\sum\limits_{k'} {\sum\limits_{m,n} {\Gamma _{in,mj} \left( {k + q,k';k' + q,k} \right)\left( {\hat G\left( {k' + q} \right)\hat \Lambda _{\mu } \left( {k' + q,k'} \right)\hat G\left( {k'} \right)} \right)_{m,n} } } , \label{eqnam5}
\end{align}
%
%
where irreducible four-point vertex $\Gamma _{in,mj} $ is expressed by the functional derivative of the self-energy matrix by the Green's function matrix:
%
%
\begin{align}
\Gamma _{in,mj}  = \frac{{\delta \Sigma _{ij} \left[ {\hat G} \right]}}{{\delta G_{mn} }}. \label{eqnam6}
\end{align}
%
%

After rather long calculation shown in Refs.\citen{bib1} and \citen{bib1a}, Eq.\ref{eq001} is rewritten as 
\begin{align}
\lambda _{\mu \nu }^{ - 2}   = 8{\mathrm{\pi }}\int_{FS} {\frac{{dS_{{\mib{k'}}} }}
{{\left( {2{\rm{\pi }}} \right)^3 \left| {v^*_{\mib{k'}}} \right|}}v^* _{\mib{k'} \mu} \left( {1 - Y\left( {{\mib{k'}};T} \right)} \right)\bar v^* _{\mib{k'}\nu } }, \label{eq1021}
\end{align}
%
where $\bar v^* _{\mib{k} }$ is the renormalized velocity of quasiparticles in the superconducting state. See Ref.\citen{bib1} for the detailed calculation of $\bar v^* _{\mib{k} }$. $ Y\left( {{\mib{k}};T} \right)$ is Yosida function:
%
%
\begin{align}
Y\left( {{\mib{k}};T} \right) =  - \int {d\varepsilon _{\mib{k}}^* \left( {\frac{{\partial f\left( {E_{\mib{k}} } \right)}}{{\partial E_{\mib{k}} }}} \right)}. \label{eq456}
\end{align}
%
%
In Eq.\ref{eq456}, $f\left( x \right) = \left( {e^{x/T}  + 1} \right)^{ - 1} $ is the Fermi distribution function and $E_{\mib{k}}  = \sqrt {\varepsilon _{\mib{k}}^{*2}  + \left(\Delta^* _{\mib{k}}  \right)^2 } $, where $\Delta^* _{\mib{k}}= z_{\mib{k}}\Delta _{\mib{k}}$ is the renormalized superconducting gap. $\Delta _{\mib{k}} =\Sigma _{a} \left( {{\mib{k}},\varepsilon  = 0} \right)$ is calculated by the analytic continuation $\mathrm{i}\varepsilon _n  \to \varepsilon  + \mathrm{i}\delta $ from the anomalous self-energy term $\Sigma _{a} \left(k \right)$. It is important that Yosida function is expressed as a universal even function of $y=\Delta^* _{\mib{k}} /T$:
%
%
\begin{align}
g\left( y \right) = \int_{ - \infty }^\infty  {e^{\sqrt {x^2  + y^2 } } \left( {1 + e^{\sqrt {x^2  + y^2 } } } \right)^{ - 2} dx}. \label{eqyosi}
\end{align}
%
%
%
%
\begin{figure}
\includegraphics[width=7cm]{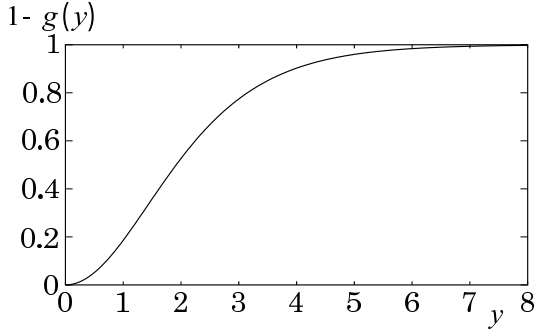}
\caption[]{The graph of $ 1-g\left( y \right) $. For $y>0$, $1-g\left( y \right)$ increases monotonically from zero at $y=0$ to unity at $y=\infty $. }
\label{yosida}
\end{figure}
%
%

The graph of $1- g\left( y \right) $ for $y>0$ is shown in Fig.\ref{yosida}. Note that $1-g\left( 0 \right)=0$ and $1-g\left(\pm \infty \right)=1$, respectively. This indicates that $1- Y\left( {{\mib{k}};T} \right)=0$ at $T=T_c$, since $\Delta^* _{\mib{k}} =0$, while $1- Y\left( {{\mib{k}};T} \right)=1$ at $T=0$, and that the increase in $y=\Delta^* _{\mib{k}} /T$ results in the increase in $\lambda^{-2}$ via Yosida function. According to Ref.\citen{bib1}, $\bar v^* _\nu  \left( {{\mib{k'}}} \right)= v^* _\nu  \left( {{\mib{k'}}} \right)$ at $T=T_c$ and $\bar v^* _\nu  \left( {{\mib{k'}}} \right)= \bar j^* _\nu  \left( {{\mib{k'}}} \right)$ at $T=0$. This means that $\bar v^* _\nu  \left( {{\mib{k'}}} \right)$ is suppressed near $T=0$ compared to that near $T=T_c$. From these facts, the inverse square of the penetration depth at $T=0$ is given by
%
%
%
\begin{align}
\left.  {\lambda _{\mu \nu } ^{ - 2} } \right|_{T = 0}  = 8{\mathrm{\pi }}\int_{FS} {\frac{{dS_{{\mib{k'}}} }}
{{\left( {2{\mathrm{\pi }}} \right)^3 \left| { v_{\mib{k'}} ^*} \right|}} v_{\mib{k'} \mu} ^* \bar j_{\mib{k'} \nu} ^*}.  \label{eq104}
\end{align}
%
%
Here, $\bar j_{\mib{k}} ^*$ is the current carried by quasiparticles in the superconducting state and $ v_{\mib{k}} ^*$ is the renormalized velocity of quasiparticles in the normal state, respectively. According to Refs.\citen{bib1} and \citen{bib1a}, $\bar j_{\mib{k}} ^*$ coincides with $ j_{\mib{k}} ^*$ due to the cancellation of the change of the velocity $\bar v_{\mib{k}} ^* - v_{\mib{k}} ^*$ and the change of the backflow effect. Therefore we use $ j_{\mib{k}} ^*$ instead of  $\bar j_{\mib{k}} ^*$ hereafter. The velocity of quasiparticle $ v_{\mib{k}} ^*$ is expressed as
%
%
\begin{align}
v_{\mib{k}}^*  = z_{\mib{k}} \frac{\partial }
{{\partial {\mib{k}}}}\left( {\varepsilon _{\mib{k}}  + \operatorname{Re} \Sigma _n^R \left( {{\mib{k}},\varepsilon =0} \right)} \right) \equiv \frac{{\partial \varepsilon _{\mib{k}}^* }}{{\partial {\mib{k}}}}. \label{eq1041}
\end{align}

After rather long calculation\cite{bib1a}, $ j_{\mib{k}} ^*$ is expressed as
\begin{align}
j_{\mib{k}}^*  = v_{\mib{k}}^*  + z_{\mib{k}} w_{\mib{k}}^R \left( 0 \right), \label{eq110x}
\end{align}
where
\begin{align}
w_{\mib{k}}^R \left( 0 \right) = u_{\mib{k}}^R \left( 0 \right)  \nonumber \\
{} + \frac{1}{N}\sum\limits_{{\mib{k'}}} {\int {\frac{{{\rm{d}}\varepsilon '}}{{2\pi }}\left[ {\coth \frac{{\varepsilon '}}{{2T}}{\mathop{\rm Im}\nolimits} V_{{\mib{k}} - {\mib{k'}}}^R \left( { - \varepsilon '} \right)\frac{{\partial {\mathop{\rm Re}\nolimits} G_{{\mib{k'}}}^R \left( {\varepsilon '} \right)}}{{\partial \varepsilon '}} + \tanh \frac{{\varepsilon '}}{{2T}}\frac{{\partial {\mathop{\rm Re}\nolimits} V_{{\mib{k}} - {\mib{k'}}}^R \left( { - \varepsilon '} \right)}}{{\partial \varepsilon '}}{\mathop{\rm Im}\nolimits} G_{{\mib{k'}}}^R \left( {\varepsilon '} \right)} \right]} z_{{\mib{k'}}} w_{\mib{k}}^R \left( 0 \right)} , \label{eq111}
\end{align}
\begin{align}
u_{\mib{k}}^R \left( 0 \right) = \frac{1}{N}\sum\limits_{{\mib{k'}}} {\int {\frac{{{\rm{d}}\varepsilon '}}{{2\pi }}\left[ {\coth \frac{{\varepsilon '}}{{2T}}{\mathop{\rm Im}\nolimits} V_{{\mib{k}} - {\mib{k'}}}^R \left( { - \varepsilon '} \right)\frac{{\partial {\mathop{\rm Re}\nolimits} G_{{\mib{k'}}}^R \left( {\varepsilon '} \right)}}{{\partial \varepsilon '}} + \tanh \frac{{\varepsilon '}}{{2T}}\frac{{\partial {\mathop{\rm Re}\nolimits} V_{{\mib{k}} - {\mib{k'}}}^R \left( { - \varepsilon '} \right)}}{{\partial \varepsilon '}}{\mathop{\rm Im}\nolimits} G_{{\mib{k'}}}^R \left( {\varepsilon '} \right)} \right]} \left( {v_{{\mib{k'}}}  - v_{\mib{k}} } \right)}  .  \label{eq112}
\end{align}
%
%
$ z_{\mib{k}}  = \left( {1 - \left. {\frac{{\partial \operatorname{Re} \Sigma _n^R \left( {{\mib{k}},\varepsilon } \right)}}
{{\partial \varepsilon }}} \right|_{\varepsilon  = 0} } \right)^{ - 1} $ is the renormalization factor. $ \Sigma _n^R \left( {{\mib{k}},\varepsilon } \right)$ and the retarded Green's function $ G_{{\mib{k}}}^R \left( {\varepsilon } \right)$ is calculated by the analytic continuation $\mathrm{i}\varepsilon _n  \to \varepsilon  + \mathrm{i}\delta $ from $\Sigma _{n} \left(k \right)$ and $ G \left(k \right)$, respectively. Similarly, $V_{{\mib{k}}}^R \left( { \varepsilon } \right)$ is derived by the analytic continuation $\mathrm{i}\omega _n  \to \varepsilon  + \mathrm{i}\delta $ from the normal vertex $ V \left(k \right)$.

Note that $ j_{\mib{k}} ^* = \left. {\bar v^* \left( {\mib{k}} \right)} \right|_{T = 0} $ is lowered compared to $ v_{\mib{k}} ^* = \left. {\bar v^* \left( {\mib{k}} \right)} \right|_{T = T_c}$.  This difference is called the vertex correction, which is considered as the backflow effect. According to Ref.\citen{bib3a}, the vertex correction is more effective, when the antiferromagnetic spin fluctuation is strong, such as in the underdoped LSCO.  The penetration depth $  {\lambda _{0 \mu \nu } ^{ - 2} }$  without including the vertex correction is expressed by
%
%
\begin{align}
\lambda _{0\mu \nu } ^{ - 2}  = 8{\rm{\pi }}\int_{FS} {\frac{{dS_{{\mib{k'}}} }}{{\left( {2{\rm{\pi }}} \right)^3 \left| {v_{{\mib{k'}}}^* } \right|}}v_{{\mib{k'}}\mu }^* \left( {1 - Y\left( {{\mib{k'}};T} \right)} \right)v_{{\mib{k'}}\nu }^* } , \label{eq00120} \\
\left.  {\lambda _{0 \mu \nu } ^{ - 2} } \right|_{T = 0}  = 8{\mathrm{\pi }}\int_{FS} {\frac{{dS_{{\mib{k'}}} }}
{{\left( {2{\mathrm{\pi }}} \right)^3 \left| { v_{\mib{k'}} ^*} \right|}} v_{\mib{k'} \mu} ^* v_{\mib{k'} \nu} ^*}.  \label{eq0012}
\end{align} 
%
%
Owing to the vertex correction, $\left.  {\lambda _{\mu \nu } ^{ - 2} } \right|_{T = 0}$ is lowered compared to $ \left.  {\lambda _{0 \mu \nu } ^{ - 2} } \right|_{T = 0}$.  Since the vertex correction is effective only near $T=0$, $\lambda _{\mu \nu } ^{ - 2}$ is suppressed near $T=0$ under the strong antiferromagnetic fluctuation.  
In order to calculate $\lambda _{0 \mu \nu } ^{ - 2}$ at finite temperature $0<T<T_c$, we have to calculate it in the superconducting state, since Eq.\ref{eq00120} contains Yosida function. We calculate $\lambda _{0 \mu \nu } ^{ - 2}$ using Eqs.\ref{eq00120}, \ref{eq456} and \ref{eq1041}, in which Green's functions and self-energy terms are calculated by Eqs.\ref{se1}-\ref{eqkai}. Although the vertex correction is not included in $\lambda _{0 \mu \nu } ^{ - 2}$, it expresses the behavior of $\lambda _{ \mu \nu } ^{ - 2}$ near $T=T_c$ correctly where the vertex correction is not effective. This fact means that we can discuss the development of $\lambda _{ \mu \nu } ^{ - 2}$ just below $T_c$ using the result of $\lambda _{0 \mu \nu } ^{ - 2}$.


\section{\label{secsec}The Pressure Dependence of the Parameters}
In this section, we discuss how the parameters in our model depend on the pressure.  We assume that the dependence of the Coulomb repulsion $U$ on the pressure is weaker than those of the hopping terms $t$ and $t'$.  On the other hand, the absolute values of $t$ and $t'$ increase under high pressure, since the lattice constant of $\kappa$-$($ET$)_2$Cu(NCS)$ _2$ decreases due to compression. Owing to the large isothermal compressibility\cite{bib12}, $\kappa$-$($ET$)_2$Cu(NCS)$ _2$ is quite sensitive to the pressure. Since the bandwidth $W$ is approximately proportional to $t$, the large value of $t$ results in small ratio of $U/W$ which indicates the weak electron correlation.

Moreover, we need to take account of the change of $t'/t$ under pressure. In Refs.\citen{bib7a} and \citen{bib7b}, $t'/t$ and the effective mass ratio for $\alpha$-pocket electron $m_{\alpha}^*/m_e$ under pressure are measured by Shubnikov-de-Haas effect in both $ \kappa$-$( d_8$ET$)_2$Cu(NCS)$ _2$ and $ \kappa$-$( h_8$ET$)_2$Cu(NCS)$ _2$. According to these results, the decrease in $T_c$, that in $m_{\alpha}^*/m_e$, and the increase in $t'/t$ are observed as the hydrostatic pressure increases. The decrease in $T_c$ is similar to the case of other organic superconductors.\cite{bib17} The decrease in the effective mass ratio is observed also by the infrared reflectivity experiment. \cite{bibIR} The stronger pressure dependence of $t'/t$, $T_c$ and $m_{\alpha}^*/m_e$ in $ \kappa$-$( d_8$ET$)_2$Cu(NCS)$ _2$\cite{bib7a} compared to that in $ \kappa$-$( h_8$ET$)_2$Cu(NCS)$ _2$\cite{bib7b} is considered to result from the different pressure medium used in the two experiments. In fact, the isotope effect for $T_c$ is small, when the same pressure medium is used. \cite{bib11} We can consider that $T_c$ under pressure depends strongly on the pressure medium.

 In $ \kappa$-$( d_8$ET$)_2$Cu(NCS)$ _2$, the ratio $t'/t$ varies from about 0.7 at $P=0$ to about 0.77 at $P=0.1$GPa. The fact that $t'/t$ increases under pressure indicates that $t'$ is more sensitive to the pressure than $t$. On the other hand, $m_{\alpha}^*/m_e$ varies from 3.5 at $P=0$ to 2.5 at $P=0.1$GPa. The change of $t'/t$ results in the change of the Fermi surface. When $t'/t$ is large, the lattice structure is near to the triangular lattice, and the Fermi surface is far from the antiferromagnetic Brillouin zone boundary as shown in Fig.\ref{fig000x}. Therefore the antiferromagnetic fluctuation becomes weaker as $t'/t$ increases with the fixed value of $U$.  From these reasons, we can say that the electron correlation and the antiferromagnetic fluctuation in $\kappa$-$($ET$)_2$Cu(NCS)$ _2$ are strong under low pressure compared with those under high pressure.  In our model, we assume that the value of $U$ is independent of the pressure and that the values of $t/U$ and $t'/t$ vary with the pressure.  From now on, we take $U$ as the unit of energy, since it is assumed to be unchanged.  In our model, $W$ and the strength of electron correlation mainly depend on $t/U$. On the other hand, the shape of the Fermi surface is changed by $t'/t$. Note that the antiferromagnetic fluctuation becomes weaker as $t'/t$ increases with the constant value of $U$.

\section{\label{sec3a} Result and Discussion }
\begin{figure}
\includegraphics[width=7.5cm]{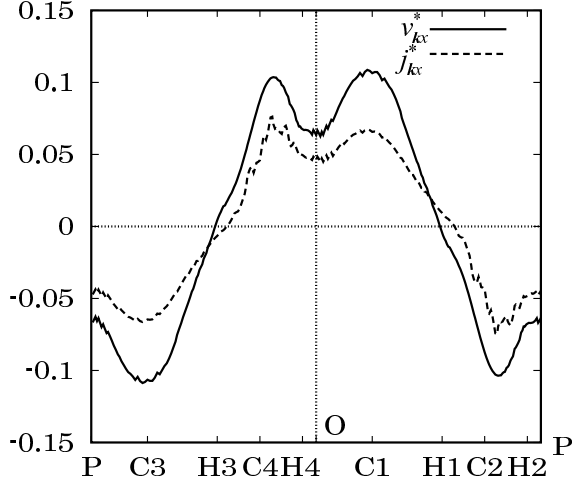}
\caption[]{The current and the renormalized velocity on the FS for $t/U=0. 20$, $t'/t=0.7$ and the temperature $T=0. 0014U$  within FLEX. The points O,P,C1-C4,H1-H4 correspond to those in Fig.\ref{fig000x}. }
\label{fig2}
\end{figure}

\begin{figure}
\includegraphics[width=7.5cm]{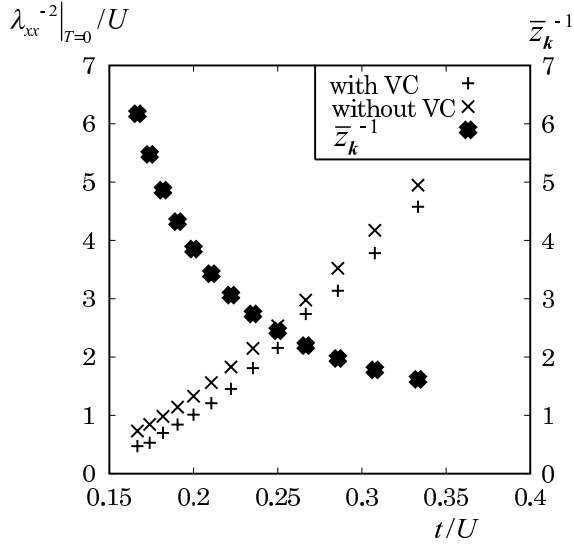}
\caption[]{$\left.  {\lambda  ^{ - 2} } \right|_{T = 0}$(with VC), $\left.  {\lambda_0  ^{ - 2} } \right|_{T = 0}$(without VC) and $\bar z_{\mib{k}}^{ - 1}$ calculated by FLEX. Low value of $t/U$ corresponds to the strong electron correlation and high value to the weak one. the ratio $t'/t$ is fixed to 0.7. The vertex correction reduces largely the value of $\lambda  ^{ - 2}$ compared to $\lambda_0  ^{ - 2}$, when the electron correlation is strong.}
\label{fig1}
\end{figure}

\begin{figure}
\includegraphics[width=7.5cm]{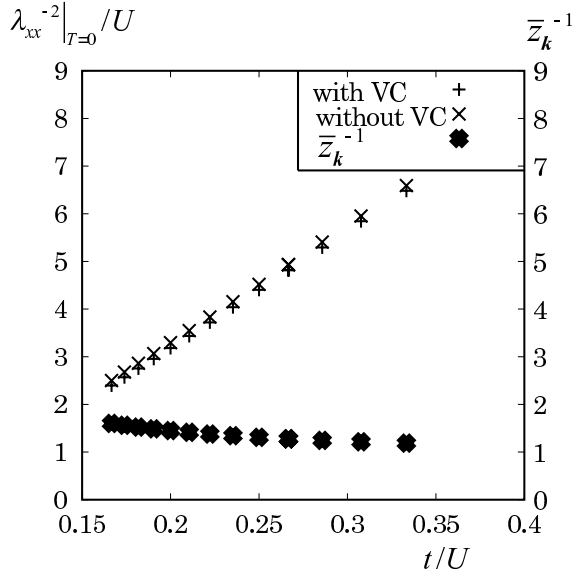}
\caption[]{$\left.  {\lambda  ^{ - 2} } \right|_{T = 0}$(with VC), $\left.  {\lambda_0  ^{ - 2} } \right|_{T = 0}$(without VC) and $\bar z_{\mib{k}}^{ - 1}$  calculated by SC-SOPT. $t'/t$ is fixed to 0.7. The reduction of $\lambda  ^{ - 2}$ by the effect of the vertex correction is quite small compared to Fig.\ref{fig1}.}
\label{fig10}
\end{figure}

\begin{figure}
\includegraphics[width=7.5cm]{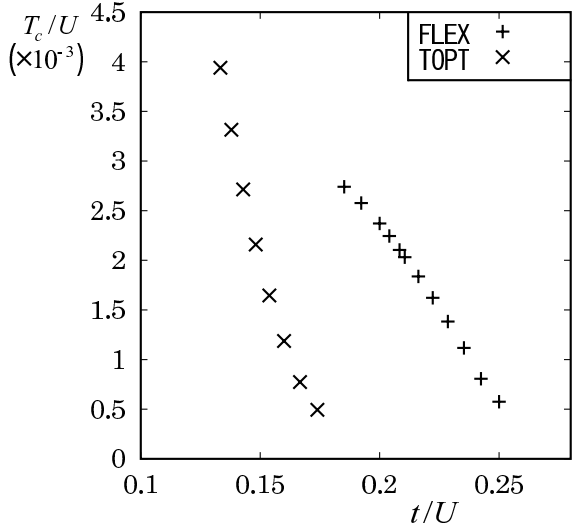}
\caption[]{The dependence of $T_c$ on $t/U$ with $t'/t=0.7$ within FLEX and TOPT. Both for FLEX and for TOPT, $T_c$ is higher when the electron correlation is strong, that is, $t/U$ is small. $T_c$ decreases more rapidly as $t/U$ increases for TOPT than for FLEX.}
\label{figtc1}
\end{figure}

\begin{figure}
\includegraphics[width=7.5cm]{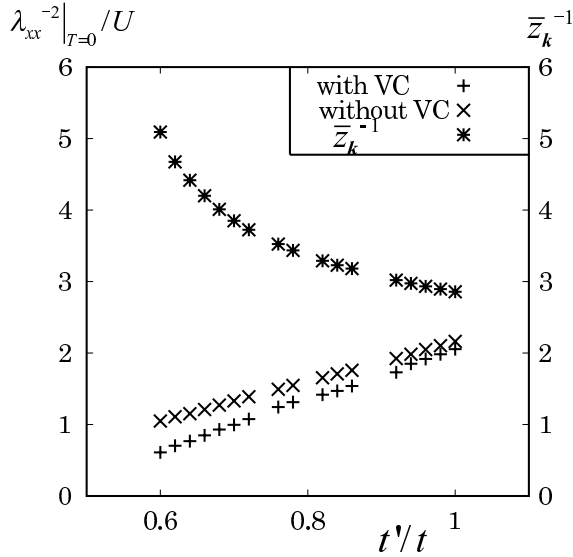}
\caption[]{The $t'/t$ dependence of  $\bar z_{\mib{k}}^{ - 1}$, $\left.  {\lambda _0^{ - 2} } \right|_{T = 0}$ and $\left.  {\lambda ^{ - 2} } \right|_{T = 0}$ with $t/U=0.200$ within FLEX. The small value of $t'/t$ corresponds to the strong antiferromagnetic spin fluctuation, while the large value to the weak fluctuation. The vertex correction affects strongly under the strong antiferromagnetic fluctuation.}
\label{figtdash}
\end{figure}

In our numerical calculation, we divide the first Brillouin zone into $128 \times 128 $ meshes in the $\mib{k}$-space and take 4096 Matsubara frequencies. In this condition, the obtained results are reliable with temperature $T$ down to approximately $0.001t$.
Within the FLEX approximation, we calculate $ v_{\mib{k} x} ^*$, $ j_{\mib{k} x} ^*$, $\left.  {\lambda _{0 } ^{ - 2} } \right|_{T = 0}$ and $\left.  {\lambda ^{ - 2} } \right|_{T = 0}$ by using Eqs.\ref{eq1041}, \ref{eq110x}-\ref{eq112}, \ref{eq0012} and \ref{eq104} respectively. The results are shown in Figs. \ref{fig2}-\ref{figtdash}. From now on, we calculate only $\lambda_{x x}^{-2}$ and refer to it as the in-plane penetration depth $\lambda_{\mu \nu}^{-2}$, since the qualitative behavior of $\lambda_{\mu \nu}^{-2}$ is similar in all directions within the $xy$-plane. We abbreviate the subscripts $xx$ and denote $\lambda_{x x}^{-2}$ as $\lambda^{-2}$ hereafter. In the calculations of $\left.  {\lambda _{0 } ^{ - 2} } \right|_{T = 0}$ and $\left.  {\lambda ^{ - 2} } \right|_{T = 0}$, the temperature at which the calculation is performed is important. In order to calculate Eqs.\ref{eq104} and \ref{eq0012} precisely, we need to obtain the values of $ v_{\mib{k} x} ^*$ and $ j_{\mib{k} x} ^*$ at $T=0$ by extrapolation of $ v_{\mib{k} x} ^*$ and $ j_{\mib{k} x} ^*$ obtained at finite temperatures. However, the extrapolation of $ j_{\mib{k} x} ^*$ is difficult, since the numerical error in the calculation increases in the low temperature region. Therefore $ v_{\mib{k} x} ^*$, $ j_{\mib{k} x} ^*$, $\left.  {\lambda _{0 } ^{ - 2} } \right|_{T = 0}$ and $\left.  {\lambda ^{ - 2} } \right|_{T = 0}$ shown in Figs.\ref{fig2}-\ref{fig10} and \ref{figtdash} are calculated with the temperature fixed to $T=0.0014U$. Although the calculations are not correct, it is enough to investigate the qualitative behavior of the vertex correction.
%
%
%
%
%
%
%
%
%
%
%
%
%
%
%
%
\begin{figure}
\includegraphics[width=7.5cm]{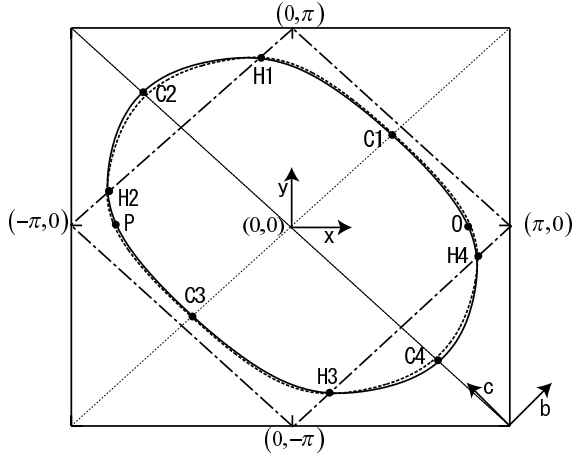}
\caption[]{The Fermi surfaces of $\kappa$-$($ET$)_2$Cu(NCS)$ _2$ with $t'/t=0.7$ (dotted line) and $0.75$ (solid line), where $t/U$ is fixed to 0.200. The shape of the Fermi surfaces is consistent with the estimation from the Shubnikov-de-Haas effect. The crossing points of the FS and the antiferromagnetic Brillouin zone boundary shown by the dash dotted line are the hot spots, which are denoted by H1-H4. On the other hand, the crossing points of the FS and the gap nodes shown by two diagonal lines are the cold spots, which are denoted by C1-C4. Note that the axes are rotated by $\pi/4$ compared to the FS in Refs.\citen{bib7a} and \citen{bib8}.}
\label{fig000x}
\end{figure}
%
%
%
\begin{figure}
\includegraphics[width=7.5cm]{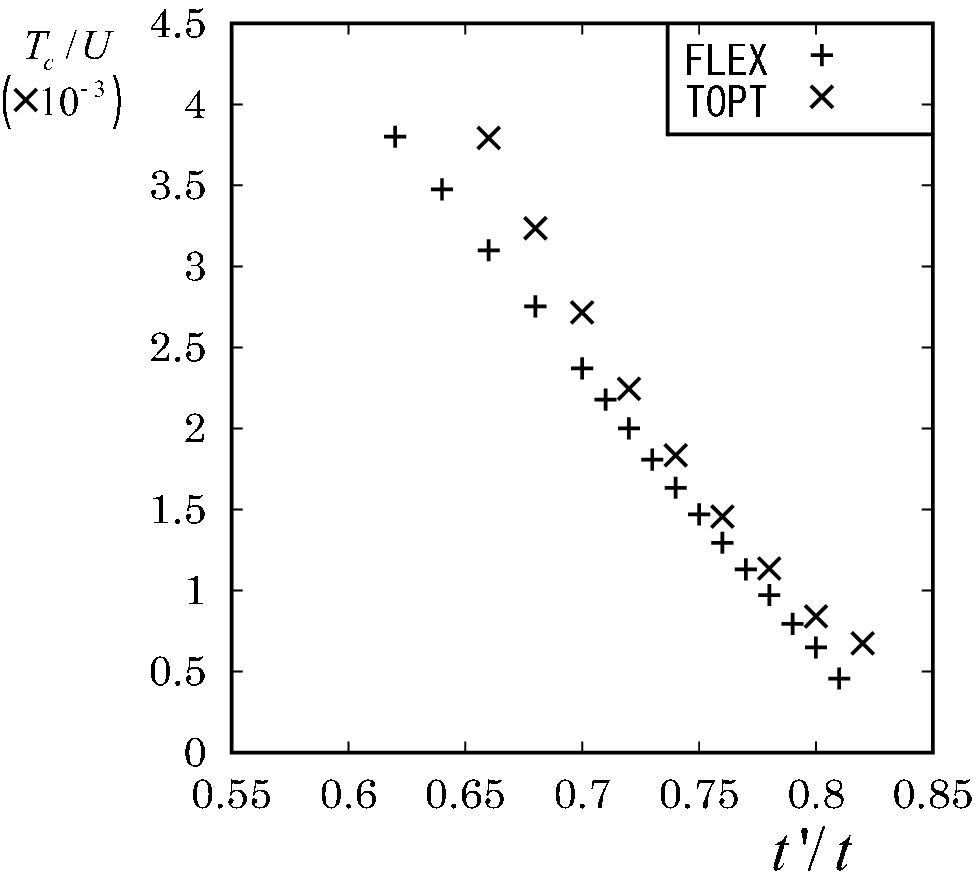}
\caption[]{The dependence of $T_c$ on $t'/t$ within FLEX and TOPT. The former is calculated with $t/U=0.200$, and the latter with $t/U=0.154$. Both for FLEX and TOPT, $T_c$ is higher under strong antiferromagnetic fluctuations, or small value of $t'/t$.}
\label{figtc2}
\end{figure}
%
%

Figure \ref{fig2} shows that $ j_{\mib{k} x} ^*$ is lowered compared to $ v_{k x} ^*$ by the vertex correction within the FLEX approximation. Note that the ratio of the vertex correction $ \left(  j_{k x } ^* - v_{k x }^* \right) /v_{k x } ^*$ is approximately constant everywhere on the Fermi surface in Fig.\ref{fig2}. This is in contrast to that in underdoped cuprates calculated in Ref.\citen{bib1}. In the cuprates,  $ \left(  j_{k x } ^* - v_{k x } \right)^*/v_{k x } ^*$ is large at hot spots compared to that at cold spots. This implies that in  $\kappa$-$($ET$)_2$Cu(NCS)$ _2$ the difference between hot spots and cold spots (see Fig.\ref{fig000x}) on the Fermi surfaces is small in contrast to the underdoped cuprates, as long as the temperature near $T=0$ is concerned. Figure \ref{fig1} shows $\bar z_{\mib{k}}^{ - 1}$, $\left.  {\lambda _0^{ - 2} } \right|_{T = 0}$, and $\left.  {\lambda ^{ - 2} } \right|_{T = 0}$ for various values of $t/U$ with $t'/t=0.7$, where $ z^{-1}=\bar z_{\mib{k}}^{ - 1}$ is the average of $ z_{\mib{k}}^{ - 1}$ over the Fermi surface. The factor $ z^{-1}$ is considered to be approximately proportional to $m_{\alpha}^*/m_e$. Figure \ref{fig10} shows that the effect of the vertex correction within SC-SOPT is quite small compared to that in the FLEX shown in Fig. \ref{fig1}. This is similar to the results in Ref.\citen{bib1a}. Figure \ref{figtc1} shows the dependence of $T_c$ on $t/U$ with $t'/t=0.7$ calculated within FLEX and within TOPT. The value of $T_c$ increases as $t/U$ decreases both for FLEX and for TOPT. For TOPT, small value of $t/U$ is required for superconductivity and $T_c$ decreases rapidly as $t/U$ increases compared to $T_c$ calculated within FLEX. These results are consistent with the calculations in Refs.\citen{biborg}, \citen{bib14} and \citen{bib15}. From Fig.\ref{fig1}, we can see that $(\left.  {\lambda _{0 } ^{ - 2} } \right|_{T = 0}-\left.  {\lambda ^{ - 2} } \right|_{T = 0})/\left. {\lambda _{0 } ^{ - 2} } \right|_{T = 0}$  increases as $t/U$ decreases. This means that the suppression of $\left.  {\lambda  ^{ - 2} } \right|_{T = 0}$ by the vertex correction  is large under the strong electron correlation, even if the shape of the Fermi surface is unchanged. Moreover, $\left.  {\lambda_0  ^{ - 2} } \right|_{T = 0}$ itself decreases as $t/U$ decreases. This is because $\left.  {\lambda_0  ^{ - 2} } \right|_{T = 0}$ is approximately proportional to $zt$, or $zW$, where $z= \left( {m^* /m} \right)^{ - 1}$ is the average of the renormalization factor on the Fermi surface. Shortly to say, the strong electron correlation suppresses $\left.  {\lambda  ^{ - 2} } \right|_{T = 0}$ through the increase of the vertex correction effect and the strong mass renormalization which corresponds to the decrease in $z$.

Figure \ref{figtdash} shows the $t'/t$ dependence of $z^{ - 1}$, $\left.  {\lambda _0^{ - 2} } \right|_{T = 0}$ and $\left.  {\lambda ^{ - 2} } \right|_{T = 0}$ with $t/U=0.200$, calculated within FLEX. On the other hand, Fig.\ref{figtc2} shows the $t'/t$ dependence of $T_c$ within FLEX and within TOPT. Both in FLEX and in TOPT, $T_c$ increases as $t'/t$ decreases and the dependence is quite similar to each other. According to Fig.\ref{figtdash}, $(\left.  {\lambda _{0 } ^{ - 2} } \right|_{T = 0}-\left.  {\lambda ^{ - 2} } \right|_{T = 0})/\left. {\lambda _{0 } ^{ - 2} } \right|_{T = 0}$ increases as $t'/t$ decreases. This means that the effect of the vertex correction in $\left.  {\lambda  ^{ - 2} } \right|_{T = 0}$ is strong under the strong antiferromagnetic fluctuation. Moreover, $\left.  {\lambda ^{ - 2} } \right|_{T = 0}$ decreases approximately in proportion to $z$ as $t'/t$ decreases. Then with similar discussion to the dependence on $t/U$, $\left.  {\lambda ^{ - 2} } \right|_{T = 0}$ is suppressed by the effect of vertex correction and by the decrease in $z$, when $t'/t $ is small. 

Under low pressure, the strong electron correlation and the strong antiferromagnetic fluctuation exist. In our model, the decrease in $t/U$ and that in $t'/t$ under low pressure indicate the strong electron correlation and the strong antiferromagnetic fluctuation, respectively. These effects result in the strong mass renormalization and the increase in the vertex correction effect. The suppression of  $\left.  {\lambda ^{ - 2} } \right|_{T = 0}$ under low pressure results from these effects. On the other hand, the increase of $T_c$ under low pressure is also explained by the strong electron correlation and the antiferromagnetic fluctuation.

Let us consider the justification of our calculation by estimating the parameters. By comparing our result for $z^{-1}$ in Figs.\ref{fig2}-\ref{figtdash} with the value of $m_{\alpha}^*/m_e$ and $t'/t$ measured in Ref.\citen{bib7a}, we can estimate the parameters in the calculation for $\kappa$-$( d_8$ET$)_2$Cu(NCS)$ _2$. For convenience, we assume $z^{-1}$ to be equivalent to $m_{\alpha}^*/m_e$ in the estimation. Judging from the values of $z^{-1}$, the FLEX approximation is more appropriate for $\kappa$-$( d_8$ET$)_2$Cu(NCS)$ _2$ than SC-SOPT. From this reason, we take account of only the FLEX approximation in the discussion of $\lambda ^{ - 2}$ hereafter. Then, we estimate $t/U=0.208$ at $P=0$ from $z^{-1}=3.5$ and $t'/t=0.7$\cite{bib7a} and $t/U=0.237$ at $P=0.1$GPa from $z^{-1}=2.5$ and $t'/t=0.77$, by using the result within FLEX. From the above estimation, we can compare the calculated $T_c$ with the measured one. Since the value of $t$ at $P=0$ is estimated\cite{bib7b} as 0.12eV $\sim$ 1400K from the polarized infrared reflectance measurements\cite{bib9}, $U$ is estimated as approximately 6700K. Using these parameters, we calculate $T_c=2.08 \times 10^{ - 3} U=14$K at $P=0$, and $T_c=2.3 \times 10^{ - 4} U=1.5$K at $P=0.1$GPa within FLEX.  The value of $T_c$ measured by experiment for  $\kappa$-$(d_8$ET$)_2$Cu(NCS)$ _2$ is approximately 10K at $P=0$ and 6K at $P=0.1$GPa\cite{bib10,bib11} although it depends strongly on the pressure medium. Considering that the calculated $T_c$ has delicate dependence on the parameters, we can say that the calculated $T_c$ is in magnitude consistent with the measured $T_c$ and that our model and calculation are justified.

\begin{figure}
\includegraphics[width=7.5cm]{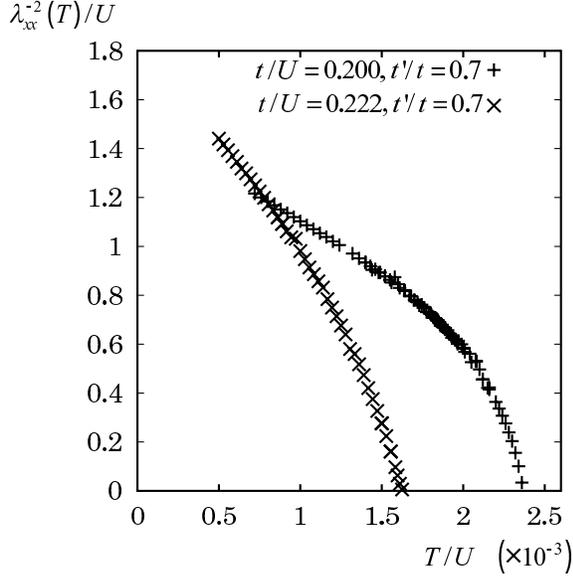}
\caption[]{${\lambda _{0 } ^{ - 2} } $-$T$ diagram for ($t/U=0.222$, $t'/t= 0.7$) and for ($ t/U=0.200$, $t'/t= 0.7$). $T_c=1.63\times 10^3U$ in the former, and $2.37\times 10^3U$ in the latter.}
\label{fig101}
\end{figure}

\begin{figure}
\includegraphics[width=7.5cm]{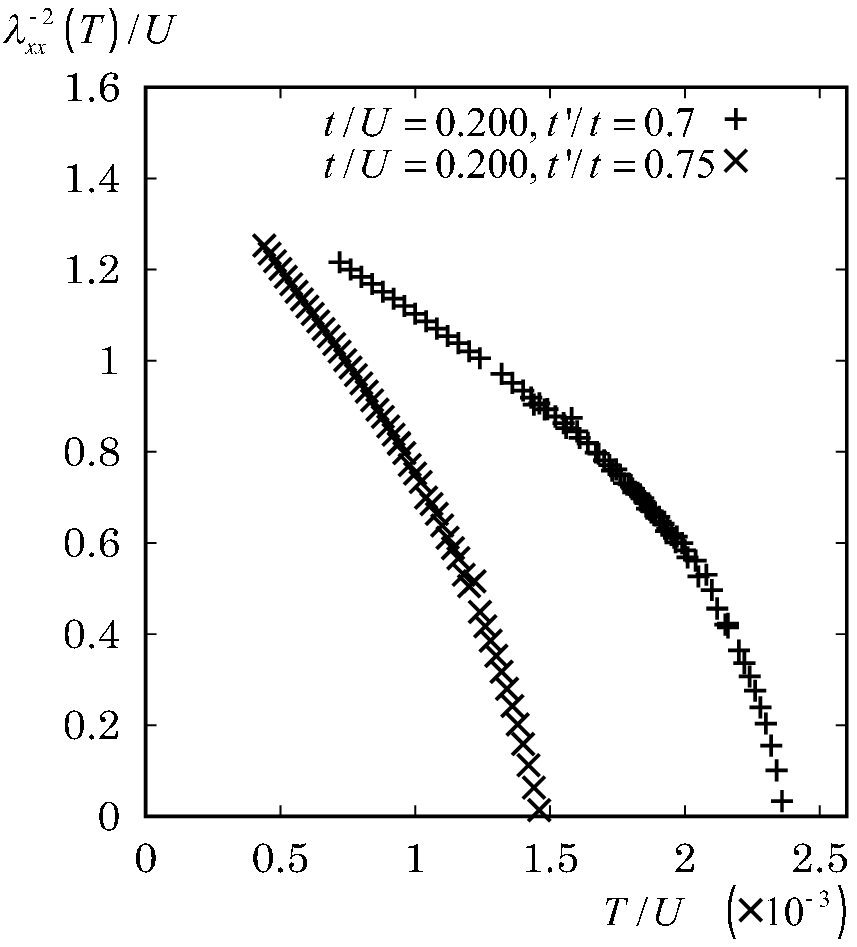}
\caption[]{${\lambda _{0 } ^{ - 2} } $-$T$ diagram for ($t/U=0.200$, $t'/t= 0.7$) and for ($ t/U=0.200$, $t'/t= 0.75$). $T_c=2.37\times 10^3U$ in the former, and $1.47\times 10^3U$ in the latter.}
\label{fig101tdash}
\end{figure}

Next, we discuss the development of $\lambda ^{ - 2} $ just below $T_c$ in $ \kappa $-$( d_8$ET$)_2$Cu(NCS)$ _2$. We calculate $\lambda _0^{ - 2} $ for $0<T<T_c$ according to Eq.\ref{eq00120}. As a matter of fact, $\lambda _0^{ - 2} $ develops rapidly just below $T_c$, when $t/U$ and $t'/t$ are small, as shown in Figs.\ref{fig101} and \ref{fig101tdash}.
Figure \ref{fig101} shows the ${\lambda_0  ^{ - 2} } $-$T$ diagram with $t/U=0.222$, $t'/t=0.7$ and with $t/U=0.200$, $t'/t=0.7$. On the other hand, Fig.\ref{fig101tdash} shows the diagram with $t/U=0.200$, $t'/t=0.7$ and with $t/U=0.200$, $t'/t=0.75$. Although the effect of vertex correction is not included in  ${\lambda_0  ^{ - 2} } $, these diagrams express qualitatively correct development of ${\lambda ^{ - 2} } $ just below $T_c$, since ${\lambda_0  ^{ - 2} } ={\lambda  ^{ - 2} } $ at $T=T_c$\cite{bib1,bib1a}. From Figs.\ref{fig101} and \ref{fig101tdash}, we can see that $\lambda _{0 } ^{ - 2}$ has a linear dependence on $T$ near $T=0$:
%
\begin{align}
\lambda _{0 } ^{ - 2}-\left. {\lambda _{0* } ^{ - 2} } \right|_{T = 0} \propto T, \label{eqlamz}
\end{align}
%
where $\left. {\lambda _{0* } ^{ - 2} } \right|_{T = 0}$ is obtained from extrapolation of $\lambda _{0 } ^{ - 2}$ calculated by Eq.\ref{eq00120} to $T=0$. The linear dependence of $\lambda _{0 } ^{ - 2}$ results from the existence of the line-node in the superconducting gap. $\left. {\lambda _{0* } ^{ - 2} } \right|_{T = 0}$ does not coincide with $\left. {\lambda _{0 } ^{ - 2} } \right|_{T = 0}$ shown in Figs.\ref{fig1} and \ref{figtdash}. This is because the results in Figs.\ref{fig1} and \ref{figtdash} are calculated with $T=0.0014U$ and are not extrapolated to $T=0$, owing to the difficulties in calculating $ j_{\mib{k} x} ^*$ just near $T=0$. Since $ v_{\mib{k} x} ^*$ can be calculated just near $T=0$, we can extrapolate  Eq.\ref{eq0012} to $T=0$. The difference between $\left. {\lambda _{0* } ^{ - 2} } \right|_{T = 0}$ and the result of extrapolation of Eq.\ref{eq0012} is less than 3\%. This indicates that the calculation of Eq.\ref{eq00120} and that of Eq.\ref{eq0012} are consistent with each other. 

\begin{figure}
\includegraphics[width=7.5cm]{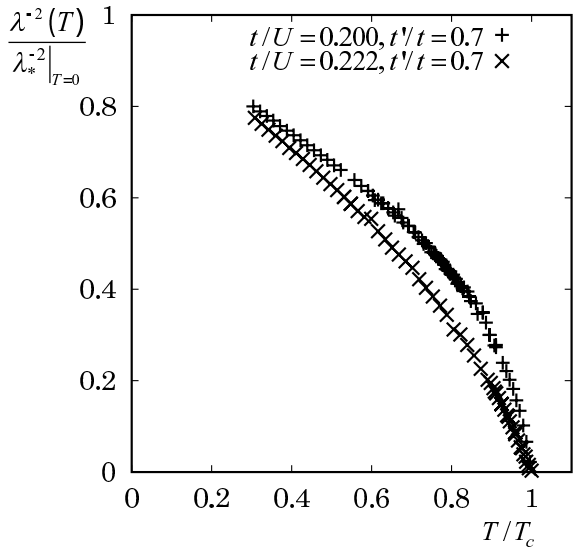}
\caption[]{${\lambda _{0 } ^{ - 2} } $-$T$ diagram for ($t/U=0.222$, $t'/t= 0.7$) and for ($ t/U=0.200$, $t'/t= 0.7$). This diagram is normalized by $T= T_c$ and $\lambda _{0 } ^{ - 2}=\left. {\lambda _{0 } ^{ - 2} } \right|_{T = 0}$, respectively. ${\lambda _{0 } ^{ - 2} } $ shows rapid development under the strong electron correlation. ($ t/U=0.200$) }
\label{fig11}
\end{figure}

\begin{figure}
\includegraphics[width=7.5cm]{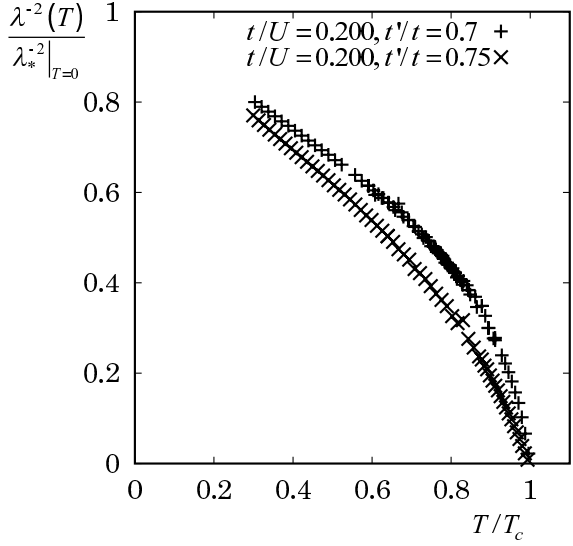}
\caption[]{${\lambda _{0 } ^{ - 2} } $-$T$ diagram for ($t/U=0.200$, $t'/t= 0.7$) and for ($ t/U=0.200$, $t'/t= 0.75$). This diagram is normalized by $T= T_c$ and $\lambda _{0 } ^{ - 2}=\left. {\lambda _{0 } ^{ - 2} } \right|_{T = 0}$, respectively. ${\lambda _{0 } ^{ - 2} } $ shows rapid development under the strong antiferromagnetic fluctuation. ($ t'/t=0.70$) }
\label{fig11tdash}
\end{figure}

For the convenience in comparison, we normalize these diagrams by $T_c$ and $\left. {\lambda _{0* } ^{ - 2} } \right|_{T = 0}$.
The results of the normalization for Figs.\ref{fig101} and \ref{fig101tdash} are shown by Figs.\ref{fig11} and \ref{fig11tdash}, respectively. In Figs.\ref{fig11} and \ref{fig11tdash}, the development of  $\lambda_0 ^{ - 2} $ just below $T_c$ with $t/U=0.200$ and $t'/t=0.7$ is more rapid than that with $t/U=0.222$ and $t'/t=0.7$, or than that with $t/U=0.200$ and $t'/t=0.75$. This indicates that the strong electron correlation and the strong antiferromagnetic fluctuation are the main origin of the rapid development of $\lambda  ^{ - 2} $ just below $ T_c$ under low pressure in Ref.\citen{bib4}.

\begin{figure}
\includegraphics[width=7.5cm]{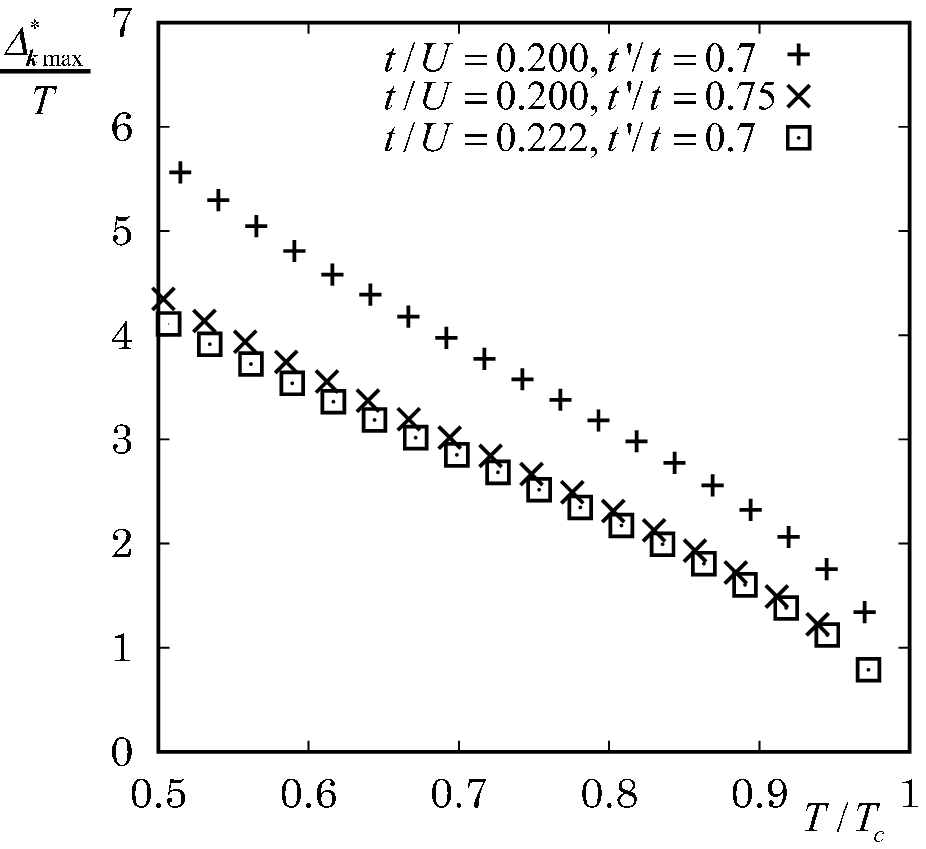}
\caption[]{The temperature dependence of $\Delta^* _{\mib{k} \mathrm{max}} /T$ with $t/U=0.200$ and $t'/t=0.700$, with $t/U=0.222$ and $t'/t=0.700$, and with $t/U=0.200$ and $t'/t=0.750$. $\Delta^* _{\mib{k} \mathrm{max}} /T$ with $t/U=0.200$ and $t'/t=0.700$ develops most rapidly.}
\label{figgap}
\end{figure}

The rapid development of $\lambda _0^{ - 2} $ just below $T_c$ implies that $\Delta^* _{\mib{k}}/T $ develops rapidly. The temperature dependence of $\Delta^* _{\mib{k} \mathrm{max}} /T$ is shown in Fig.\ref{figgap}, where $\Delta^* _{\mib{k} \mathrm{max}}$ is the maximum of $\Delta^* _{\mib{k}} $ on the Fermi surface. In Fig.\ref{figgap}, $\Delta^* _{\mib{k} \mathrm{max}} /T$ develops rapidly with $t/U=0.200$ and $t'/t=0.700$, compared to that with $t/U=0.222$ and $t'/t=0.700$, or to that with $t/U=0.200$ and $t'/t=0.750$. This indicates that the strong antiferromagnetic fluctuation and the strong electron correlation cause the rapid development of $\Delta^* _{\mib{k} \mathrm{max}} /T$, which result in the rapid development of $\lambda _0^{ - 2} $. Figures \ref{fig11}, \ref{fig11tdash}, and \ref{figgap} justify the conclusion that the increases in $\Delta^* _{\mib{k}}/T $ is the dominant cause of the development of $\lambda _0^{ - 2} $.
The pseudogap effect is also considered to be the cause of the rapid development of $\Delta^* _{\mib{k}} $ just below $T_c$, although it can not be calculated in the framework of this paper. The pseudogap arises from the superconducting fluctuation as discussed by Jujo, \etal\cite{bibpg} In fact, the pseudogap effect in $\kappa$-$( d_8$ET$)_2$Cu(NCS)$ _2$ is observed up to $T=45$K in STM spectroscopy at ambient pressure.\cite{bibSTM1,bibSTM2} Although similar measurements under pressure are not performed, it is natural to consider that the pseudogap effect is weak or disappears under pressure owing to the weak electron correlation and the weak antiferromagnetic fluctuation. If the pseudogap exists above $ T_c$, then the superconducting gap $\Delta^* _{\mib{k}} $ develops rapidly when the superconducting state is realized at $T=T_c$ as shown in Ref.\citen{bibpseudo}.

%
%
%
%
%
%
%
%
%
%
%
%
%
%
%
%
%
%
%
%
Last, we discuss the uniaxial pressure dependence of $T_c$. We are not able to calculate the effect of the interlayer pressure applied parallel to $a$-axis, since we adopt Q2D band structure and ignore the interlayer hopping term. Nevertheless, we can presume that the three-dimensionality is strong under the interlayer pressure and it is generally understood that $T_c$ is suppressed under strong three-dimensionality as calculated in Refs.\citen{bib3d} and \citen{bibfu}. For the $b$-axis pressure, the behavior of $T_c$ is easily explained. Under the $b$-axis pressure, both $t$ and $t'$ increase. Since the increase in $t$ is small and that in $t'$ is large under the $b$-axis pressure compared to those under the hydrostatic pressure, it is sure that both $t'/t$ and $t/U$ increase under the $b$-axis pressure. Therefore we can say that the $b$-axis pressure suppresses $T_c$ similarly to the hydrostatic pressure from the discussion above. 

Then we discuss the behavior of $T_c$ under the $c$-axis pressure. Since the $c$-axis is perpendicular to the direction of the hopping term $t'$, it is natural to assume that $t'$ is constant and that $t$ increases under the $c$-axis pressure. Moreover, $U$ is assumed to be constant with increasing $c$-axis pressure. These assumptions indicate that $t/U$ increases and $t'/t$ decreases. The former corresponds to the suppression of the electron correlations and the latter to the deformation of the Fermi surface, which approaches the antiferromagnetic Brillouin zone boundary. These two effects are competitive to each other for $T_c$: the former decreases $T_c$ and the latter increases it.

\begin{figure}
\includegraphics[width=7.5cm]{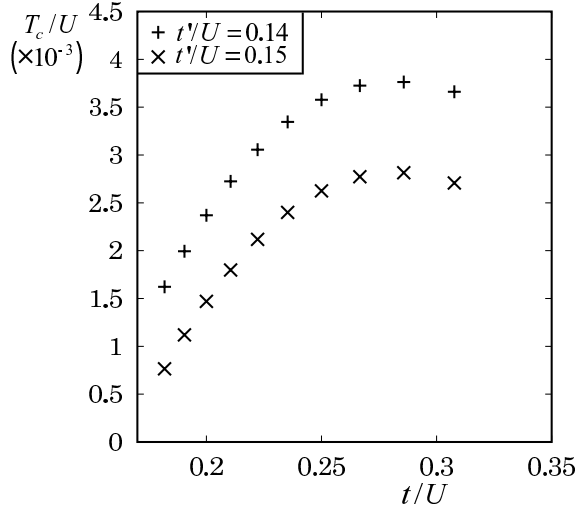}
\caption[]{The $c$-axis pressure dependence of $T_c$ calculated within FLEX. The value of $t'/U$ is fixed to 0.14 (which has higher $T_c$) and 0.15(lower). $T_c$ tends to increase rapidly up to $t/U$ becomes 0.28.}
\label{figuni1}
\end{figure}

\begin{figure}
\includegraphics[width=7.5cm]{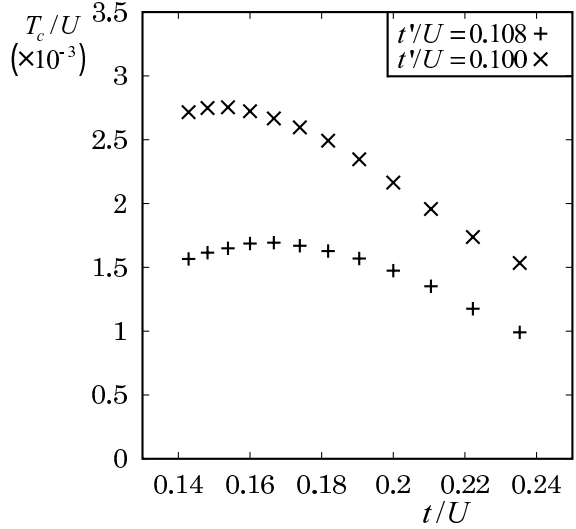}
\caption[]{The $c$-axis pressure dependence of $T_c$ calculated within TOPT. The value of $t'/U$ is fixed to 0.100 (which has higher $T_c$) and 0.108(lower). $T_c$ slightly increases up to $t/U=0.16$ and then decreases rapidly.}
\label{figuni2}
\end{figure}

Figures \ref{figuni1} and \ref{figuni2} show the behavior of $T_c$ under the $c$-axis pressure calculated within FLEX and TOPT, respectively. For FLEX, $T_c$ tends to increase rapidly up to $t/U=0.28$ and the decrease over $t/U=0.28$ is rather slow. The value of $t/U=0.28$ corresponds to rather higher pressure than 1kbar, considering that the increase of $t$ under the uniaxial pressure is smaller than that under the hydrostatic pressure. As a conclusion, the value of $T_c$ within FLEX increases as the $c$-axis pressure increases, since the deformation of the Fermi surface affects strongly compared to the suppression of the electron correlation. This behavior of $T_c$ is consistent with the results in Refs.\citen{bib18} and \citen{bib19} below 1kbar. This fact implies that the antiferromagnetic fluctuation is strong under the weak $c$-axis pressure. On the other hand, $T_c$ calculated by TOPT increases slightly up to $t/U=0.16$ and then decreases rather rapidly. The difference between FLEX and TOPT is mainly due to the $t/U$ dependences of $T_c$ shown in Fig.\ref{figtc1}. $T_c$ by TOPT depends rather strongly on $t/U$ than that by FLEX, while the $t'/t$ dependences differ little. Although it is quite difficult to estimate the value of $t/U$ for TOPT, we estimate $t/U$ at ambient pressure to be from 0.15 to 0.16 judging from the value of $T_c \sim 11$K and $t \sim 1400$K. Therefore, $T_c$ calculated within TOPT decreases under the $c$-axis pressure, since the suppression of the electron correlation is dominant over the deformation of the Fermi surface. This behavior of $T_c$ is consistent with the experimental results\cite{bib18,bib19} on $T_c$ under the $c$-axis pressure above 1kbar. It implies that the antiferromagnetic fluctuation is weak under the high $c$-axis pressure. As a conclusion, we can explain the behavior of $T_c$ under the $c$-axis pressure by considering two competitive effects: the suppression of the electron correlation and the approach of the Fermi surface to the antiferromagnetic Brillouin zone boundary. In the low $c$-axis pressure region, the latter effect is dominant and FLEX approximation is valid. On the other hand, in the high pressure region the former effect is dominant and TOPT is valid.
%
%
%
%
%
%
%
%

\section{\label{sec4} Conclusion}
Under low pressure, the development of $\lambda ^{ - 2} $ just below $T_c$ in $\kappa$-$($ET$)_2$Cu(NCS)$ _2$ is rapid and $\lambda ^{ - 2} $ near $T=0$ is suppressed, when they are compared to those under high pressure.  These effects, added to the increase of $T_c$, result from the strong electron correlation and the strong antiferromagnetic fluctuation under low pressure. This is due to the decrease in $t/U$ and that in $t'/t$. In cuprates, the Fermi surface changes with the change of doping, and the antiferromagnetic spin fluctuation is enhanced in underdoped cuprates, while the renormalization factor $z$ and the bandwidth depend weakly on the doping.\cite{bib1,bib1a} Then the suppression of $\lambda  ^{ - 2} $ near $T=0$ in underdoped cuprates are mainly due to the vertex correction. On the other hand, the explanation of the pressure dependence of $\lambda^{-2}$ in  $\kappa$-$($ET$)_2$Cu(NCS)$ _2$ is rather complicated. Under low pressure, the narrow bandwidth $W$ and the strong antiferromagnetic fluctuation are observed. The former effect results in the strong electron correlation and the decrease in $z$, while the latter enhances the vertex correction. In short, the decrease in $z$ and the effect of the vertex correction are all the origins of the suppression of $\lambda  ^{ - 2} $ at $T=0$.

The rapid development of $\lambda  ^{ - 2} $ just below $T_c$ under low pressure in $\kappa$-$( $ET$)_2$Cu(NCS)$ _2$ is explained by the rapid development of the superconducting gap. The superconducting gap develops rapidly just below $T_c$ under low pressure, since the strong antiferromagnetic fluctuation and the strong electron correlation exist. Moreover, the pseudogap effect is also considered to cause the rapid development of the superconducting gap, although it is not studied in this paper. This is similar to the case of cuprates. In cuprates, the strong pseudogap effect is observed in the underdoped region where the antiferromagnetic fluctuation is strong, and $\lambda  ^{ - 2} $ develops rapidly just below $T_c$ compared to that in the overdoped region. 

 As a conclusion, we have explained various experimental results on $\kappa $-$( $ET$)_2$Cu(NCS)$ _2$ under pressure, \textit{e.g.} $\lambda  ^{ - 2} $ near $T=0$, the development of $\lambda  ^{ - 2} $ just below $T_c$, the superconducting transition temperature under the hydrostatic pressure and that under the uniaxial pressure. In the explanation, we adopted Hubbard model and assumed that the electron correlation and the antiferromagnetic fluctuation are suppressed under pressure. Then the obtained results are consistent with the measurements, which indicates that our assumption is appropriate.

The numerical calculation was performed by sx5 in Yukawa Institute Computer Facility, Kyoto University.

\end{document}